\documentclass[prb,twocolumn,showpacs,superscriptaddress]{revtex4}
\usepackage[latin1]{inputenc}
\usepackage{graphicx}
\usepackage{amssymb}
\usepackage{amsmath}
\usepackage{xspace}
\usepackage{dcolumn}
\usepackage{bm}
\usepackage{times}

\newfont{\tensy}{cmsy10}
\newcommand{\chem}[1]{{$\fontdimen16\tensy=3.0pt
    \fontdimen17\tensy=3.0pt \mathrm{#1}$}}

\renewcommand{\Im}[0]{\text{Im}\,}

\newcommand{\ie}[0]{i.e.\@\xspace}
\newcommand{\eg}[0]{e.g.\@\xspace}

\newcommand{\rmi}{\text{i}}

\newcommand{\om}[0]{\omega}
\newcommand{\Ep}{\varepsilon_\text{p}}
\newcommand{\EF}{E_\text{F}}
\newcommand{\kF}{k_\text{F}}
\newcommand{\nag}{\phantom{\dag}}
\newcommand{\on}{\hat{n}}

\newcommand{\las}[0]{\langle}
\newcommand{\ras}[0]{\rangle}

\renewcommand{\tilde}[1]{\widetilde{#1}}

\begin{document}

\title{Spectral signatures of the Luttinger liquid to charge-density-wave transition}
\author{M. Hohenadler}\email{hohenadler@itp.tu-graz.ac.at}
\affiliation{%
Institute for Theoretical and Computational Physics, TU Graz,
8010 Graz, Austria}
\author{G. Wellein}
\affiliation{%
Computing Center, University Erlangen, 91058 Erlangen,
Germany}
\author{A. R. Bishop}
\affiliation{%
Theoretical Division and Center for Nonlinear Studies, Los Alamos
National Laboratory, Los Alamos, New Mexico 87545, USA}
\author{A. Alvermann}
\affiliation{%
Institute for Physics, Ernst-Moritz-Arndt University Greifswald, 17487
Greifswald, Germany}
\author{H. Fehske}
\affiliation{%
Institute for Physics, Ernst-Moritz-Arndt University Greifswald, 17487
Greifswald, Germany}

\begin{abstract}
  Electron-- and phonon spectral functions of the one-dimensional,
  spinless-fermion Holstein model at half filling are calculated 
  in the four distinct regimes of the phase diagram, 
  corresponding to an attractive or repulsive Luttinger
  liquid at weak electron-phonon coupling, and a band-- or polaronic insulator
  at strong coupling. The results obtained by means of kernel polynomial 
  and systematic cluster approaches reveal substantially different physics 
  in these regimes and further indicate that the size of the 
  phonon frequency significantly affects the nature of the 
  quantum Peierls phase transition, the latter being either of the soft-mode or
  central-peak type. The generic features observed are relevant to several
  classes of low-dimensional materials.
\end{abstract} 

 
\maketitle

\section{Introduction}

Low-dimensional materials like halogen-bridged transition metal complexes,
ferroelectric perovskites, conjugated polymers, or organic charge-transfer
salts are very susceptible to structural distortions driven by
electron-phonon (EP) interaction. At commensurate band fillings these systems
might undergo a Peierls or charge-density-wave (CDW) instability accompanied
by a dimerization of the lattice but, unlike in conventional metals, both
quantum lattice fluctuations and strong electronic correlations are
important.\cite{TNYS90}  The challenge of understanding the related
metal-insulator transition, especially in the strong-EP coupling regime and
in conjunction with strong electronic correlations, has renewed the interest
in models of interacting electrons and phonons.\cite{BMH98,MHB02,Mat05,CSC05}

In this work, we compare in detail the spectral signatures of four physically
distinct regimes in the phase diagram of the one-dimensional (1D) Holstein
model of spinless fermions (HMSF) (Fig.~\ref{fig:pd-hm}). To this end, we
calculate the single-particle spectral functions of electrons and phonons, using exact
diagonalization (ED) in combination with cluster perturbation theory
(CPT)\cite{SPP02} and the kernel polynomial method (KPM).\cite{WWAF06}
Besides the influence of the degree of phonon retardation on the
nature of the Peierls transition, the wave-number dependence of the spectral functions
missed in previous dynamical mean-field theory studies\cite{MHB02} turns out
to be crucial for the 1D case considered here.

The paper is organized as follows. The model is introduced
and its phase diagram reviewed in Sec.~\ref{sec:model}. In
Sec.~\ref{sec:methods} we present the methods used, and results are discussed
in Sec.~\ref{sec:results}. Finally, we summarize in Sec.~\ref{sec:conclusions}.

\section{Model}\label{sec:model}

The Hamiltonian
\begin{equation}\label{eq:model:H}
  H
  =
  -t \sum_{\las i,j\ras} c^\dag_i c^{\nag}_j
  +\om_0\sum_i b^\dag_i b^{\nag}_i
  -g\om_0 \sum_i \on_i (b^\dag_i + b^{\nag}_i)
  \,,
\end{equation}
is particularly rewarding to study a number of interesting general phenomena
because it exhibits at half filling, as the EP coupling increases, a
zero-temperature quantum phase transition from a metallic Luttinger liquid
(LL) to an insulating Peierls state.  In Eq.~\eqref{eq:model:H}, $c^\dag_i$
($b^\dag_i$) creates an electron (dispersionless phonon of energy $\om_0$) at
site $i$ of a 1D lattice with $N$ sites.  The first term describes hopping
processes ($\propto t$) between neighboring sites $\las i,j\ras$, the second
term gives the elastic and kinetic lattice energy, and the third term
accounts for the local coupling ($\propto g$) between the lattice
displacement $\hat{x}_i=b^\dag_i+b^{\nag}_i$ and the electron density
$\on_i=c^\dag_i c^{\nag}_i$, all in the limit of infinite on-site (Hubbard)
Coulomb repulsion, \ie, the site occupation numbers are
$n_i=0,1$. Important parameters to be used here are the adiabaticity
ratio $\om_0/t$ and the dimensionless coupling $\lambda=\Ep/2t$, where
$\Ep=g^2\om_0$ is the polaron binding energy.

\begin{figure}
  \includegraphics[width=0.475\textwidth]{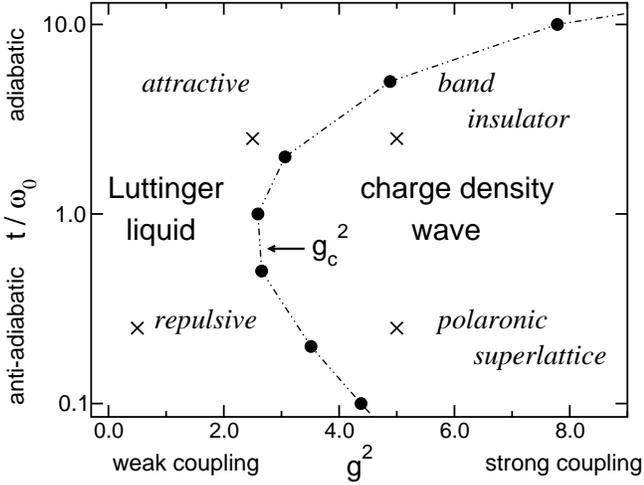}
  \caption{\label{fig:pd-hm}DMRG ground-state phase diagram
    of the 1D half-filled HMSF (cf. Refs.~\onlinecite{BMH98,FHW00}).  In the
    adiabatic limit $\om_0\to 0$ the critical dimensionless coupling constant
    $\lambda_\text{c}$ converges to zero.  For $\om_0>0$, due
    to quantum phonon fluctuations, there exists a finite critical coupling
    $g_\text{c}(\om_0)$. For $\om_0\to\infty$, the
    model exhibits a Kosterlitz-Thouless phase transition near
    $g_\text{c}^*=g_\text{c}(\om_0\to\infty)$.\cite{BMH98} Crosses
    indicate parameter sets used below.}
\end{figure}

Despite its simplicity, the HMSF is not exactly solvable and a wide range of
analytical and numerical methods have been applied to map out the
ground-state phase diagram in the $g$\,--\,$\om_0$ plane (see references
in Ref.~\onlinecite{SHBWF05}).  At present the probably most precise phase
boundary is obtained by exact diagonalization and density matrix renormalization group
(DMRG) techniques.\cite{BMH98,FHW00} More recent large-scale DMRG
calculations supplemented by a finite-size analysis have proved that at low
(high) phonon frequencies the metallic LL phase is characterised by an
attractive (repulsive) interaction.\cite{FWHWBB05} But also above
$g_c(\om_0)$, where long-range CDW order sets in, there exist two physically
distinct regimes, which can be classified, \eg, by their different optical
response,\cite{FHW00} either as a band insulator in the adiabatic regime
$\om_0/t\ll 1$ or as a polaronic superlattice in the limit of large phonon
frequencies $\om_0/t\gg 1$ (see Fig.~\ref{fig:pd-hm}).  Including the spin
degrees of freedom and a finite Hubbard interaction allows for additional
quantum phase transitions between Peierls and Mott insulating
phases.\cite{Mat05} The half-filled HMSF captures the relevant physics of
the more general quarter-filled Hubbard-Holstein model\cite{MHB02,Mat05} in
the regime of large Hubbard repulsion $U\gg t$, often realized in experiment,
where on-site bipolaron formation is suppressed.

\begin{figure*}
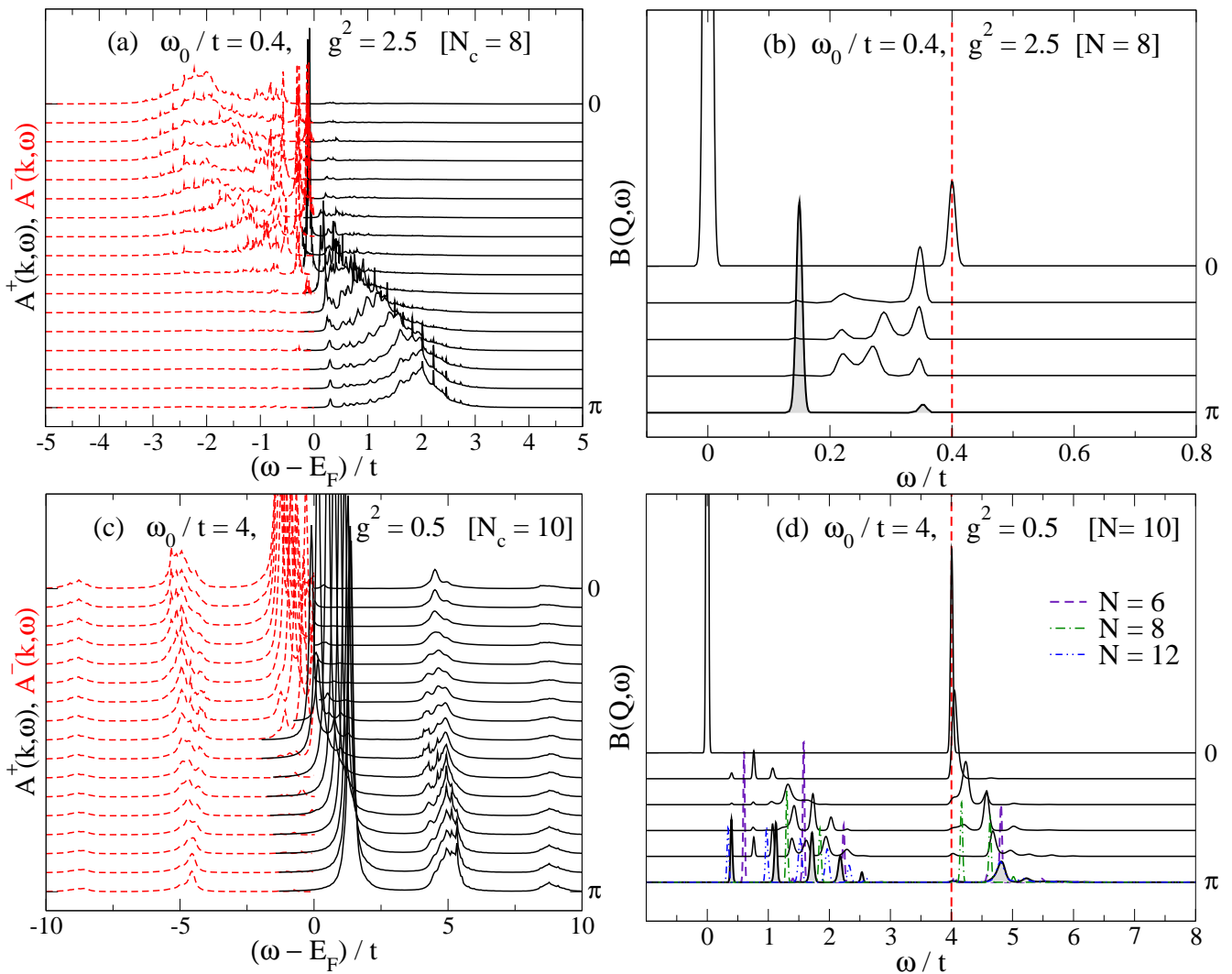

  \includegraphics[height=0.39\textwidth]{Akw_w0.4_Ep1.0.eps}
  \includegraphics[height=0.39\textwidth]{Bqw_w0.4_Ep1.0_ED.eps}\\
  \includegraphics[height=0.39\textwidth]{Akw_w4.0_Ep2.0.eps}
  \includegraphics[height=0.39\textwidth]{Bqw_w4.0_Ep2.0_ED.eps} 
  \caption{\label{fig:ll} (Color online) Left panel [(a),(c)]: Spectral
    density for photoemission [$A^-(k,\om)$; red dashed lines] and
    inverse photoemission [$A^+(k,\om)$; black solid lines] from CPT in the
    metallic LL phase. Energies are measured relative to the Fermi level $\EF$.
    Right panel [(b),(d)]: Corresponding exact phonon spectral function 
    $B(Q,\om)$ for the allowed wave-numbers $Q$ of $N$-site clusters.
    Electron and phonon spectra have been calculated in the 
    adiabatic [attractive LL; (a),(b)] and anti-adiabatic 
    [repulsive LL; (c),(d)] weak-to-intermediate EP coupling regimes. Dashed
    vertical lines in (b),~(d) indicate the bare phonon frequency.
  }
\end{figure*}

\section{Methods}\label{sec:methods}

\begin{figure*}
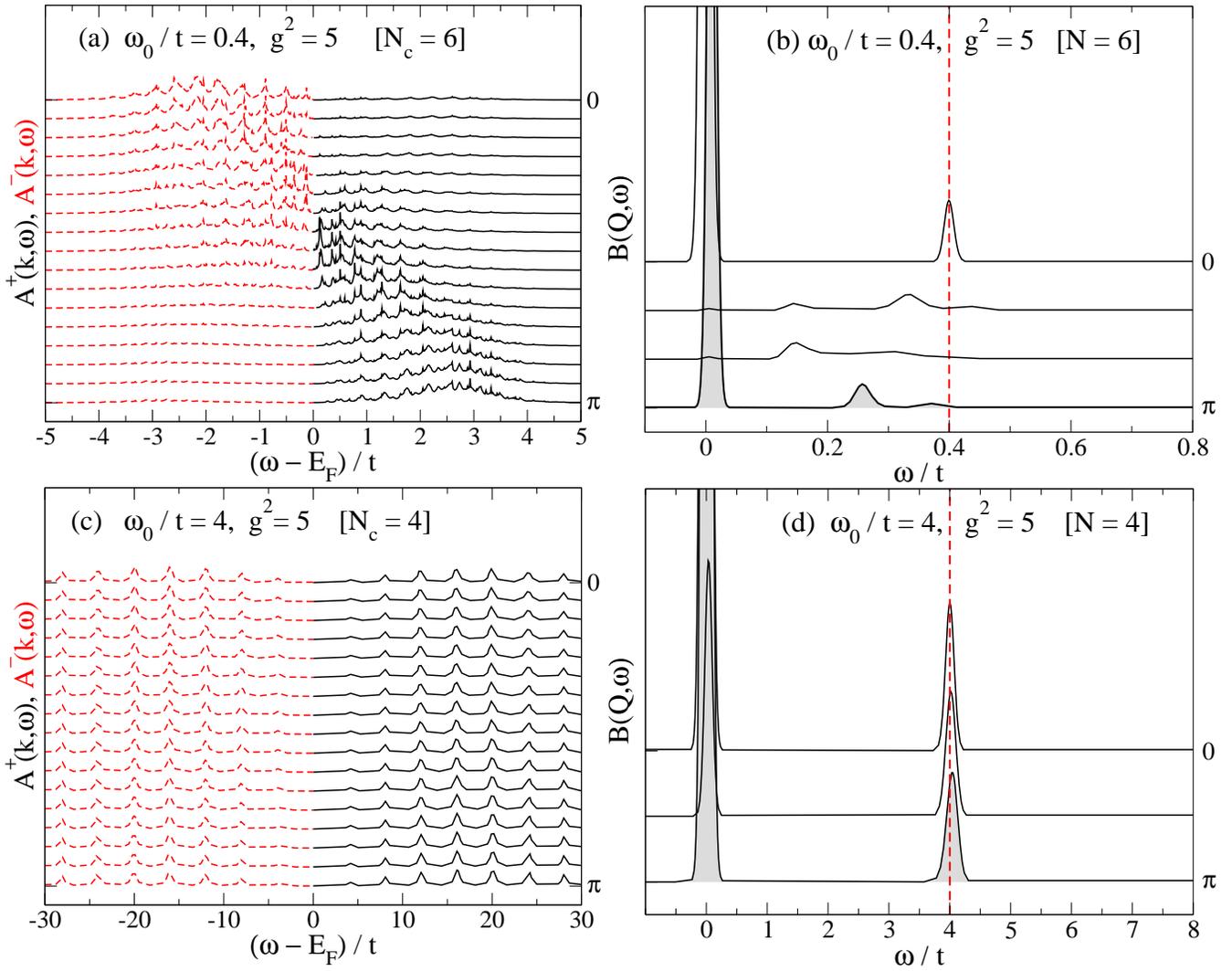
 
  \includegraphics[height=0.39\textwidth]{Akw_w0.4_Ep2.0.eps}
  \includegraphics[height=0.39\textwidth]{Bqw_w0.4_Ep2.0_ED.eps}\\
  \includegraphics[height=0.39\textwidth]{Akw_w4.0_Ep20.0.eps}
  \includegraphics[height=0.39\textwidth]{Bqw_w4.0_Ep20.0_ED.eps}
  \caption{\label{fig:cdw} 
    (Color online) As in Fig.~\ref{fig:ll}, but for parameters in the
    insulating CDW phase, and for the case of a band insulator [(a),(b)] and
    a polaronic superlattice [(c),(d)]. 
  }
\end{figure*}

The $T=0$ electron spectral function is related to the one-electron Green
function via 
\begin{equation}
  A(k,\om) 
  = 
  -\frac{1}{\pi}
  \Im\,G(k,\om) 
  = A^+(k,\om)+A^-(k,\om)
  \,,
\end{equation}
where
\begin{equation}\label{aspekt}
A^{\pm}(k,\om)=
-\frac{1}{\pi}\Im\!\!\lim_{\eta\to 0^+}
\las \psi_0 | c_k^\mp \frac{1}{\om +\rmi\eta\mp H} c_k^\pm |\psi_0\ras\,,
\end{equation}
$c_k^-=c_k^{}$, $c_k^+=c_k^\dag$, and $|\psi_0\rangle$ is the ground state of
the HMSF with $N_\text{e}=N/2$ electrons. $A^-(k,\om)$ [$A^+(k,\om)$]
describes (inverse) photoemission of an (injected) electron with momentum $k$
and energy $\om$.

Relying on approximation-free numerical approaches (here ED) we can calculate
exact Green functions for finite systems only. To obtain an approximation of
$G(k,\om)$ for the infinite lattice, we can exploit CPT.\cite{SPP02} To
this end, we divide the infinite system into identical clusters of
$N_\text{c}$ sites each, and determine the electron cluster Green function
$G^{(c)}_{ij}(\om)$ for all non-equivalent pairs of sites
$i,j=1,\dots,N_\mathrm{c}$ by the KPM (for details see Ref.~\onlinecite{WWAF06}).  The
phonon Hilbert space is truncated such that the resulting error of the
spectra is negligible ($<10^{-4}$), and we have used 1024 Chebyshev moments.
In a second step, the Green function $G(k,\om)$ for the
infinite lattice is obtained from the first-order result\cite{SPP02}
\begin{equation}
  \tilde{G}_{ij}(k,\om) = \left[ \frac{G^{(c)}(\om)}{1-\tilde{t}(k)\,
  G^{(c)}(\om)} \right]_{ij}
\end{equation}
of a strong-coupling expansion in the inter-cluster hopping operator
$\tilde{t}(k)$ as
\begin{equation}
  G(k,\om) 
  = 
  \frac{1}{N_\text{c}} \sum_{i,j=1}^{N_c} \tilde{G}_{ij}(k,\om)
  e^{-\rmi k\cdot(r_i-r_j)} \,.
\end{equation}

The $T=0$ phonon spectral function is defined as 
\begin{equation}
B(q,\om)=-\frac{1}{\pi}\Im D(q,\om)
\end{equation}
with
\begin{equation}\label{eq:phononSF}
  D(q,\om) 
  = 
  \lim_{\eta\to 0^+}\ 
  \las \psi_0 | \hat{x}_{q}
  \frac{1}{\om + \rmi\eta - H}
  \hat{x}_{-q} |\psi_0\ras
\end{equation}
for $\om\geq0$ and $\hat{x}_{q}= N^{-1/2} \sum_j \hat{x}_j e^{-\rmi
r_j\cdot q}$. For the HMSF~(\ref{eq:model:H}), $B(q,\om)$ is
symmetric in $q$, and we have a dispersionless bare propagator $D_0(q,\om) =
2\om_0/(\om^2-\om_0^2)$. EP interaction will renormalize the phonon
frequency, whereby  $D(q,\om)$ attains a $q$-dependence. Concerning
notation, we shall reserve the capital $Q$ for allowed wavenumbers of
finite clusters.

\section{Results}\label{sec:results}

The crosses in the phase diagram (Fig.~\ref{fig:pd-hm}) mark the four
parameter sets to be used in the sequel.

Figures~\ref{fig:ll}(a) and~(c) show
results for the CPT electron spectral functions $A^\pm(k,\om)$ in the
\textit{metallic region}. In the adiabatic (attractive LL) regime
[Fig.~\ref{fig:ll}(a)], we find a rather
pronounced peak at the Fermi level $\EF$ but nevertheless the system is not a
Fermi liquid.  The non-universal LL parameters (charge velocity and
interaction coefficient) can be determined from a DMRG finite-size
scaling.\cite{FWHWBB05}  At $k=\kF$, very little spectral weight is
contained in the incoherent part of the spectrum. By contrast, away from the
Fermi momentum, almost all of the spectral weight resides in the incoherent
part, whose maximum follows quite closely the free dispersion $-2t\cos k$
[notice the symmetry of $A^-(k<\kF)$ and $A^+(k>\kF)$].  As these states are
accessible only via (multi--) phonon excitations, the width of the incoherent
band is proportional to $g^2$. 

The spectrum in the anti-adiabatic (repulsive
LL) regime, Fig.~\ref{fig:ll}(c), looks significantly different. Due to the
larger coupling $\lambda=\Ep/2t=1$ (with respect to the bare
electronic bandwidth), as compared to $\lambda=0.5$ in Fig.~\ref{fig:ll}(a),
the carriers are more strongly renormalized. This is reflected in the reduced
bandwidth $W\approx3t$, with the spectral weight being more evenly
distributed than in the adiabatic case. The fact that $\om_0>W$ gives rise to
multiple, non-dispersive side-bands at energies $\EF\pm n\om_0$.

In Figs.~\ref{fig:ll}(b) and~(d) we present the corresponding phonon spectral
functions, as obtained from ED. Again we find pronounced differences between
the adiabatic [Fig.~\ref{fig:ll}(b)] and the anti-adiabatic
[Fig.~\ref{fig:ll}(d)] cases, revealing the distinct nature of the Peierls
instability---driven by $\lambda$ or $g$---for small and large $\om_0/t$.

For $\om_0/t=0.4$ [Fig.~\ref{fig:ll}(c)], we observe a peak at $\om=0$
originating from the homogeneous ($Q=0$) shift of the electronic level for
$\lambda>0$, as well as a signal located at the bare phonon energy $\om_0$.
More importantly, the phonon excitations gradually soften near the zone
boundary ($Q=\pi$) already in the LL phase, a behavior characteristic of the
Peierls transition traditionally regarded as a displacive phase transition.

For large $\om_0/t$ [Fig.~\ref{fig:ll}(d)], we observe two main absorption
features. The first excitation near $\om_0$ becomes even harder with
increasing wave number, in contrast to the phonon softening in the adiabatic
regime. The second branch emerging from $\omega=0$ ($Q=0$) resembles the two-spinon
continuum~\cite{Fadeev} of the XXZ model (onto which the spinless Holstein model may be
mapped in the anti-adiabatic strong-coupling limit\cite{BMH98}), \ie, it can be traced
back to phonon-signatures of the corresponding electronic excitations.

We now turn to the \textit{insulating state} of the HMSF
(Fig.~\ref{fig:cdw}). Due to the Peierls distortion, and the formation of
long-range CDW order for $N\to\infty$, a gap opens at $\EF$. In the adiabatic
regime, the transition is from a LL to a traditional Peierls band insulator,
whereas in the anti-adiabatic regime, the charge carriers first undergo a
cross-over to small-polarons (polaronic metal), which then (upon increasing
the coupling further) order to form a polaronic superlattice. 

From Fig.~\ref{fig:cdw}(a) we see that the gap in the single-particle spectrum is
still quite small. This is partly due to the fact that CPT in the form used
here does not fully take into account long-range order.  At $\kF$, the
strongest signatures lie just above and below $\EF$, with the incoherent part
again being very small. Away from $\kF$, the incoherent band is significantly
broadened (again almost $\propto g^2$ at $k=\pi$), and reflects the Poisson
distribution of the phonons in the ground state. Nevertheless, we can still
detect the dispersion of the split electronic band.

In contrast to the
adiabatic case of Fig.~\ref{fig:cdw}(a), Fig.~\ref{fig:cdw}(c) shows a clear
gap for all $k$. The polaronic charge carriers---in addition to forming a
superlattice---are quasi-localized, leading to dispersionless excitations
near energies $\EF\pm n\om_0$.

\begin{figure}[t]
  \includegraphics[height=0.215\textwidth]{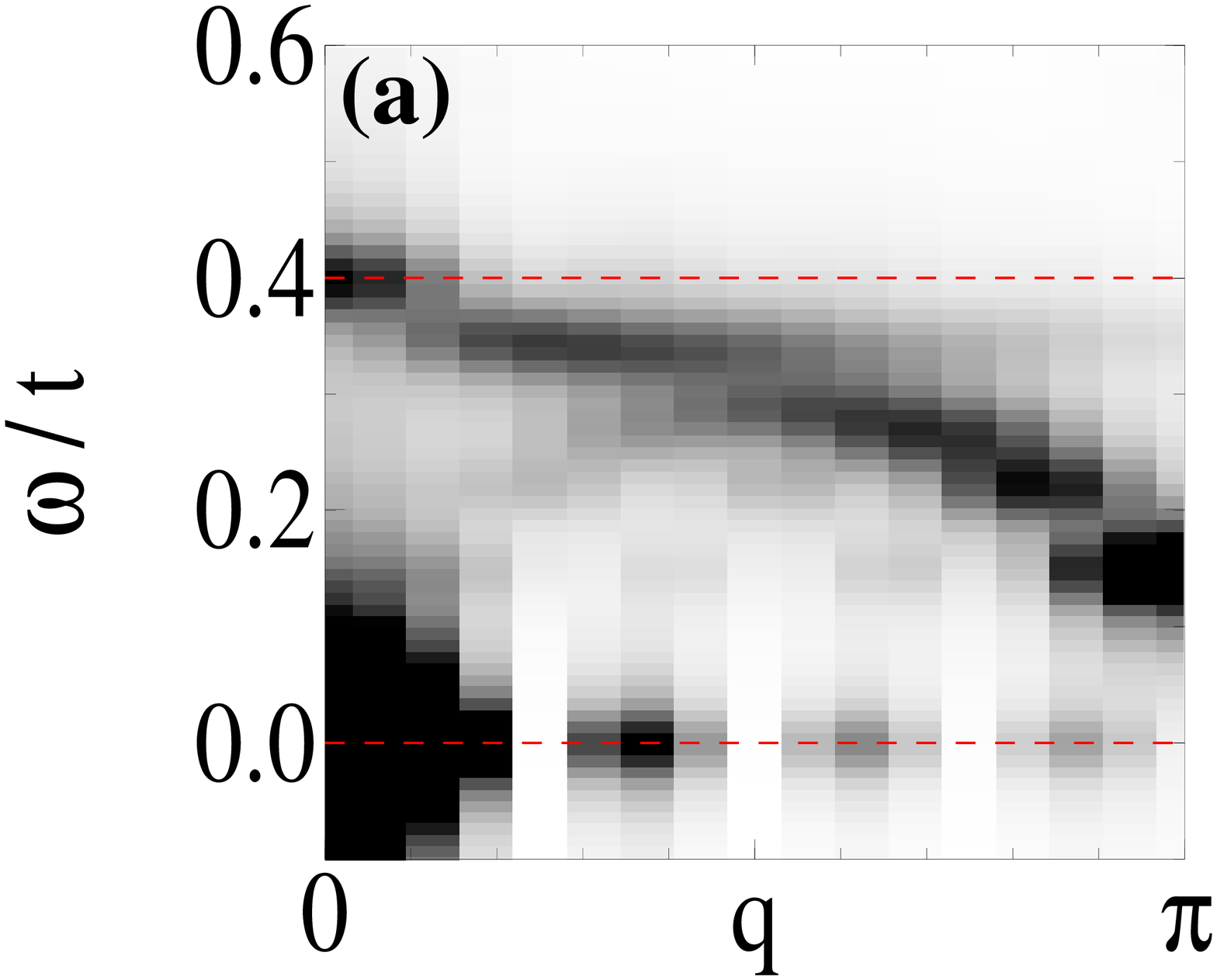}%
  \includegraphics[height=0.215\textwidth]{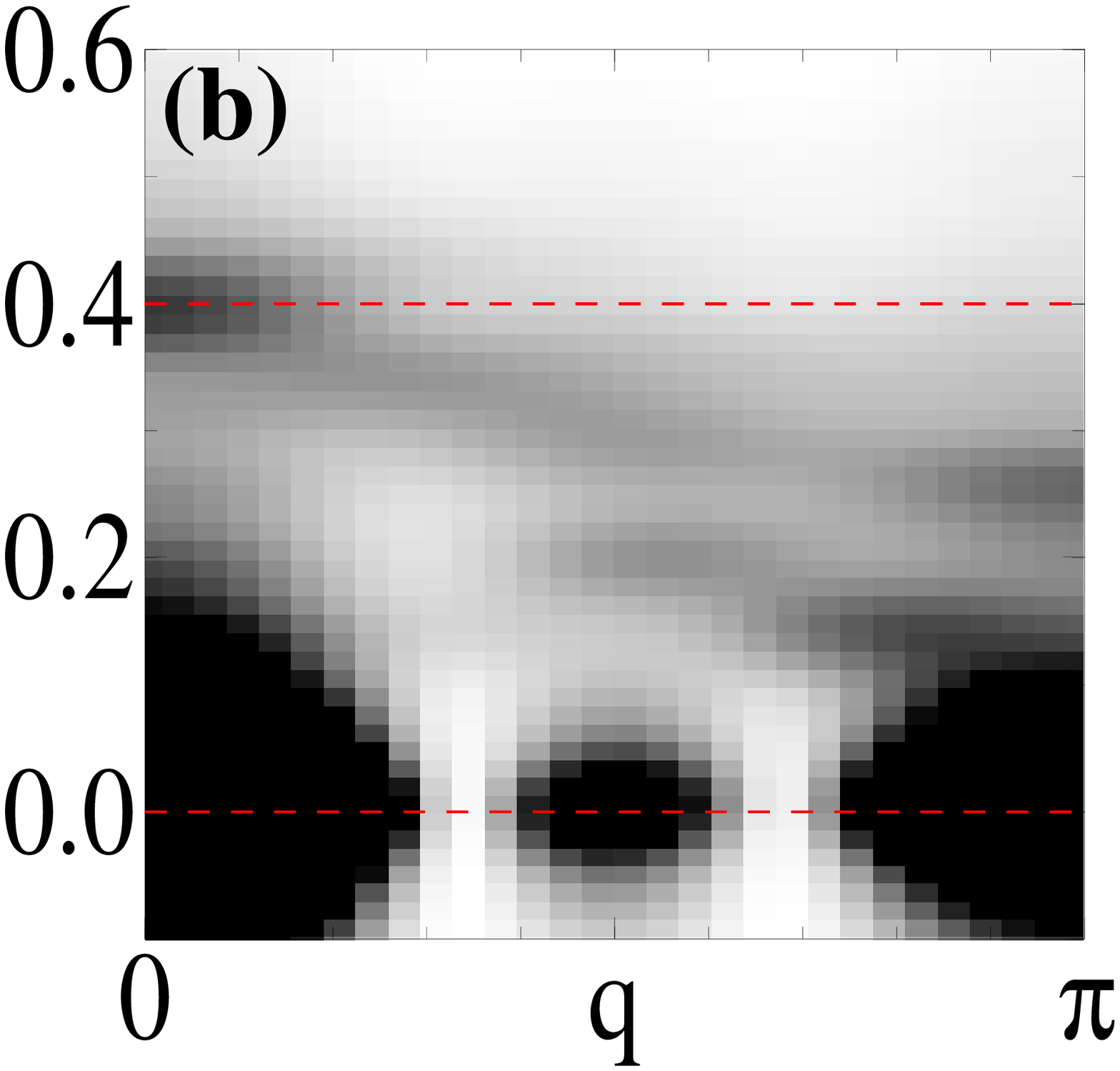}\\
  \includegraphics[height=0.215\textwidth]{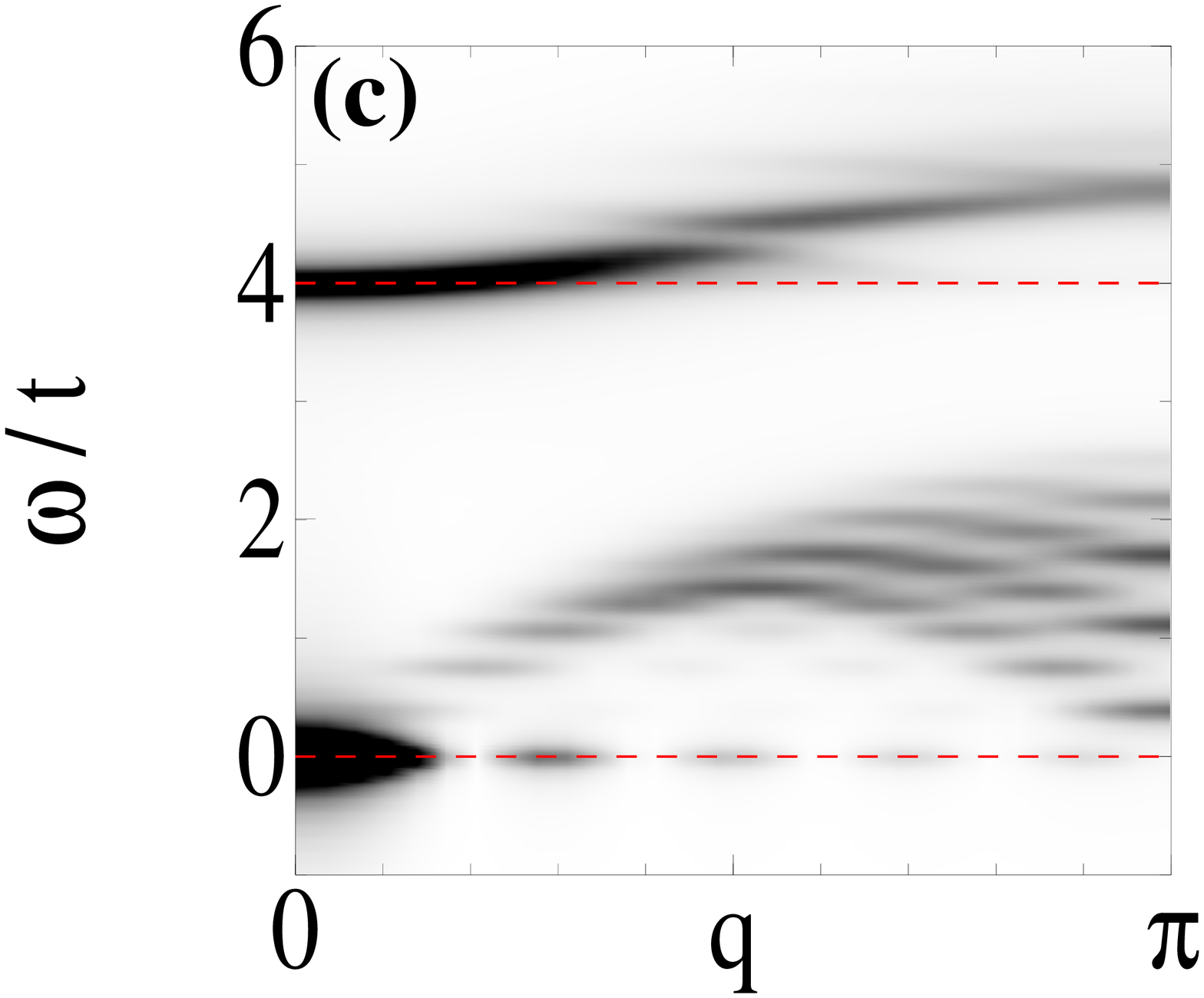}%
  \includegraphics[height=0.215\textwidth]{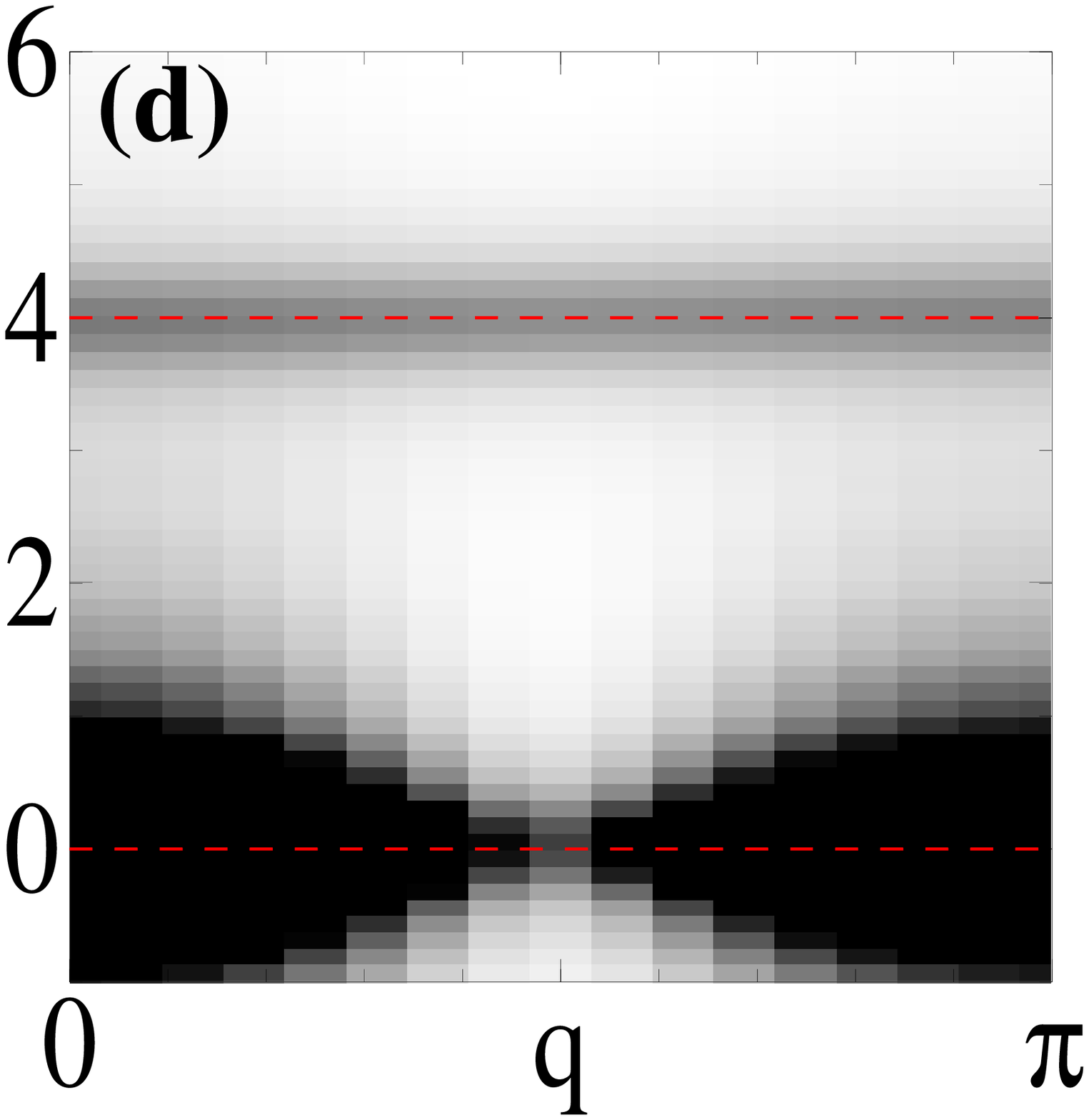}%
  \caption{\label{fig:bqw_den} (Color online) Density plots of 
    the (CPT) phonon spectral function $B(q,\om)$, where panels (a),
    (b), (c) and (d) correspond to the attractive LL ($\om_0/t=0.4$, 
    $g^2=2.5$), band insulator ($\om_0/t=0.4$, 
    $g^2=5$), repulsive LL ($\om_0/t=4$, 
    $g^2=0.5$), and polaron superlattice regimes
    ($\om_0/t=4$, $g^2=5$), respectively.}
\end{figure}

The phonon spectra for the
insulating phase are shown in Figs.~\ref{fig:cdw}(b),(d). In the adiabatic case of
Fig.~\ref{fig:cdw}(b), the zone boundary phonon has almost completely
softened.  In the limit $N\to\infty$, we expect a perfect degeneracy of
$Q=0,\pi$ for $g>g_\text{c}$, and a macroscopic population of the soft phonon
mode.  For $\om_0/t=4$ [Fig.~\ref{fig:cdw}(d)], there exist two completely
flat signatures at $\om=0$ and $\om=\om_0$ [cf. Fig.~\ref{fig:bqw_den}(d)],
the former having very large spectral weight due to the large ``phonon
content'' of the polaron band.

Similar to the results for $A(k,\om)$, it is desirable to obtain results for
the phonon spectrum of the infinite system. However, in deriving a CPT-like
approach to the phonon Green function $D(q,\om)$, it turns out that the
first-order correction in $\tilde{t}$ vanishes identically, since the
electron number per cluster is a conserved quantity. Therefore, CPT 
results in a simple Fourier transformation of the cluster
Green function (again obtained by the KPM),
\begin{equation}
  D(q,\om)
  =
  \frac{1}{N_\mathrm{c}}
  \sum_{i,j=1}^{N_\mathrm{c}} D^{(\text{c})}_{ij}(\om)
  e^{-\rmi q\cdot(r_i-r_j)}
  \,. 
\end{equation}
Notice that this result still becomes exact
for the free Green function $D_0(q,\om)$, and in the limit of strong coupling
and large phonon frequency, when $t$ is a small parameter. As the calculation
of the second-order contribution is extremely demanding we have restricted
ourselves to this expansion to illustrate the $q$-dependence
of $B(q,\om)$ in Fig.~\ref{fig:bqw_den}.

The density plots in Fig.~\ref{fig:bqw_den} clearly summarize the differences
between the adiabatic and anti-adiabatic regimes, and between the LL and CDW
phases of the HMSF.  In Fig.~\ref{fig:bqw_den}(a) we see the renormalized
phonon dispersion $\tilde{\om}(q)$, which softens with increasing EP
coupling, leading to a degeneracy of excitations at $Q=0,\pi$ at $g_\text{c}$
[see panel (b)].  The strong zero-energy absorption feature at $\pi/2$ is an
artifact of the small cluster size and the open boundary conditions used in
the CPT scheme [cf. Fig.~\ref{fig:cdw}(b)]. Above the Peierls transition we
find---in agreement with recent Monte Carlo simulations\cite{CSC05}---that
the soft $Q=\pi$ phonon mode splits into two branches with the upper one
hardening as the EP coupling increases further.

Quite differently, in the
anti-adiabatic case, we observe two phonon signatures for all $g>0$.  In the
LL phase the bare phonon mode hardens, whereas a second mode becomes strongly
over-damped near $Q=\pi$ [Fig.~\ref{fig:bqw_den}(c)]. Finally,
Fig.~\ref{fig:bqw_den}(d) reveals a dispersionless signal at $\om=\om_0$, as
well as the flat polaron band at $\om\approx0$ for the polaronic CDW state.
Thus, with increasing phonon frequency, we find a cross-over from a soft-mode
(displacive) to a central-peak-like (order-disorder-type) phase transition,
similar to the analysis of the spin-Peierls transition motivated by
\chem{CuGeO_3}.\cite{FHW00}

\section{Summary}\label{sec:conclusions}

By presenting highly-reliable numerical results for the
electron- and phonon spectral function, we have identified four physically
distinct regions in the phase diagram of the one-dimensional spinless
Holstein model, characterized by important generic features such as
attractive or repulsive Luttinger liquid behavior and phonon softening, which are highly
relevant for low-dimensional condensed matter systems.

\begin{acknowledgments}
  
This work was supported by the FWF project No.~P15834, the DFG through
SPP1073, KONWIHR and HPC Europa.  Work at Los Alamos is performed under the
auspices of USDOE.  We would like to thank M.~Aichhorn, E.~Jeckelmann,
J.~Loos, and A.~Wei{\ss}e for useful discussion, and acknowledge generous
computer time granted by the HLRN Berlin, LRZ Munich and HLR Stuttgart.

\end{acknowledgments}


\end{document}